# On the Statistical Modeling and Analysis of Repairable Systems

**Bo Henry Lindqvist**

*Abstract.* We review basic modeling approaches for failure and maintenance data from repairable systems. In particular we consider imperfect repair models, defined in terms of virtual age processes, and the trend-renewal process which extends the nonhomogeneous Poisson process and the renewal process. In the case where several systems of the same kind are observed, we show how observed covariates and unobserved heterogeneity can be included in the models. We also consider various approaches to trend testing. Modern reliability data bases usually contain information on the type of failure, the type of maintenance and so forth in addition to the failure times themselves. Basing our work on recent literature we present a framework where the observed events are modeled as marked point processes, with marks labeling the types of events. Throughout the paper the emphasis is more on modeling than on statistical inference.

## 1. INTRODUCTION

According to a commonly used definition of a repairable system [5], this is a system which, after failing to perform one or more of its functions satisfactorily, can be restored to fully satisfactory performance by a method other than replacement of the entire system. For the present paper and following recent literature on the subject, we suggest extending this definition to include the possibility of additional maintenance actions which aim at servicing the system for better performance. We shall refer to this as preventive maintenance (PM), where one may further distinguish between condition-based PM and planned PM. The former type of maintenance is due when the system exhibits inferior performance, while the latter is performed at predetermined points in time. In this presentation we will consider some aspects of condition-based PM, while the planned PM will be briefly touched on in the concluding remarks.

*Bo H. Lindqvist is Professor, Department of Mathematical Sciences, Norwegian University of Science and Technology, Trondheim, Norway e-mail: bo@math.ntnu.no.*

Traditionally, the literature on repairable systems is concerned with modeling failure times, with point process theory being the main tool. The most commonly used models for the failure process of a repairable system are renewal processes (RP), including the homogeneous Poisson processes (HPP) and nonhomogeneous Poisson processes (NHPP). While such models often are sufficient for simple reliability studies, the need for more complex models has of course emerged.

There is currently a rapidly increasing literature concerning modeling and analysis of recurrent events, with a wide range of applications, including reliability analysis of repairable systems, which is the present topic. In a recent review paper, Cook and Lawless [14] presented several examples from medical studies where models and methods for recurrent events are appropriate. The review paper by Peña [55] gave examples from both medical and reliability studies. The scope of our paper is biased toward reliability applications, although most of the models considered have a wider applicability. We will, in particular, consider models which incorporate effects of different kinds of repair and maintenance, and with the possibility of handling several failure causes, for example.

In a review paper like this, it is of course impossible to cover all models or methods which have been suggested in the literature. Our aim is rather to emphasize some important ideas, and in this respect there will be a clear bias toward work in the direction of our own interests and in work by ourselves and collaborators. Throughout the paper the emphasis will be more on modeling than on statistical inference. In addition we will try to give some historical perspectives on the theory and practice related to repairable systems, again not necessarily complete and possibly biased by our own views.

One of the first comprehensive treatments of statistical methods for recurrent events with reliability emphasis is the talk by David R. Cox, read before the Royal Statistical Society in London in March 1955 and published in [17]. Cox touched a large number of topics, most of them motivated from the clothing industry. Topics of particular importance for reliability applications were trend testing, testing whether a failure process is a Poisson process, autocorrelated time gaps, doubly stochastic Poisson processes, heterogeneity between systems, correlations between different types of events, mean repair times, availability of service and so forth. Many results from the paper are contained in the subsequent book by Cox and Lewis [19], which still is a very useful and much cited source on the subject.

Another early contribution to the study of repairable systems is the heavily cited 1963 paper by Proschan [58], "Theoretical explanation of observed decreasing failure rate." This paper is particularly important since it led to the awareness that proper analysis of recurrent events is an important part of reliability theory. In particular it is one of the first treatments of heterogeneity in the theory of repairable systems.

What seems to be the first book devoted solely to repairable systems reliability was published by Ascher and Feingold [5] in 1984. For a long time this was the main reference for repairable systems and it is still a major source. The subtitle of the book is *Modeling, Inference, Misconceptions and Their Causes*. The authors were complaining that reliability researchers and practitioners did not recognize the crucial difference between the statistical treatment of repairable systems and nonrepairable components. They demonstrated by simple examples how conclusions from data may be very wrong if times between failures are treated as i.i.d. if there is a trend in them.

Data from repairable systems are usually given as ordered failure times $T_1, T_2, \ldots$ with data coming from a single system or from several systems of the same kind. The implicit assumption is usually that the system is repaired and put into new operation immediately after the failure. This restriction, disregarding repair times, is not serious if one is interested in modeling and estimation of the probability mechanisms behind failure occurrences. It is, moreover, justified if the time scale is taken to be operation time, number of cycles, number of kilometers run and so forth. We will impose this restriction in this paper, and we will therefore not cover important topics such as availability and unavailability of systems, where the standard tool is to use alternating renewal processes with operation periods alternating with repair periods (see, e.g., [59], Chapter 7).

A common recipe for analysis of data from a repairable system is as follows. First, apply a test for trend in the interfailure times $X_i = T_i - T_{i-1}$. If no significant trend is found, then use a RP as a model, in which case the well established statistical tools for analysis of i.i.d. observations can be used. Otherwise, use a NHPP model, which handles trend through specification of an intensity function $\lambda(t)$. For example, a deteriorating system will then correspond to an increasing function $\lambda(t)$, while an improving system will correspond to a decreasing $\lambda(t)$. A homogeneous Poisson process, HPP($\lambda$), corresponds to a constant intensity $\lambda(t) \equiv \lambda$ and is at the same time a renewal process with exponentially distributed interfailure times.

A RP model is also called a perfect repair model, since the system is as good as new after a failure. On the other hand, a NHPP model corresponds to what is called minimal repairs, meaning that the system after repair is only as good as it was immediately before the failure. Lindqvist, Elvebakk and Heggland [48] represent the problem of distinguishing between the two "extreme" kinds of repair as corresponding to the first "dimension" of a repairable system description in the form of a so-called model cube (Figure 3). The second dimension is the appearance of trend or no trend in interfailure times. This particular aspect of system behavior has traditionally received much attention in reliability theory and is resolved by considering trend tests. Finally, the third dimension corresponds to the existence of unobserved heterogeneity between systems. This problem is of course relevant only when several systems of the same kind are observed. There is currently a large and increasing interest in the modeling of heterogeneity, usually known as frailties in the survival analysis literature. To some extent, heterogeneity may have been much overlooked in reliability studies, but there are important exceptions in the literature.

Several classes of models have in turn been suggested for cases not covered by the "extreme" models RP and NHPP. These include the so-called imperfect repair models. The idea is that after a repair the "virtual" age of the unit is not necessarily reduced to 0, such as for a perfect repair, nor is it the same as before the repair, such as for a minimal repair. Instead, the virtual age is reduced by a certain amount that depends on the type of repair. We review the basic properties of such models and we will see how the concept of virtual age can be generalized to more than one dimension.

Another class of alternatives to NHPP and RP models, which includes these models, is the so-called trend-renewal process (TRP). This model is a generalization of Berman's modulated gamma process [9] and has been extensively studied in [48]. In the present paper we will use TRP models and their extensions as our basic framework to illustrate some main issues on modeling and statistical analysis of data from repairable systems. The TRP is particularly suitable to illustrate the already mentioned three dimensions of repairable systems.

Modern reliability data bases usually contain more information than just the failure times. For example, there may be information on the times of preventive maintenance (PM), identity of a failed component, type of failure, type of repair, cost of replacement and so forth. Thus we shall more generally assume that observations from repairable systems are represented as marked point processes where the marks label the types of events. For example, the marks may be of two kinds, corresponding to the type of maintenance, repair or PM. We review some recent literature in this direction with the aim of extending the theory of repairable systems to a competing risks setting.

In addition to information on types of events, the data bases may contain covariates that represent environmental conditions, measures of various forms of load and usage stress, and so forth. Such covariates could be constant or are possibly varying with time. Regression models for repairable systems are useful for obtaining better understanding of the underlying failure and PM mechanisms, or for predicting the behavior of new items.

The outline of the paper is as follows. The basic notation and definitions used are given in Section 2, including the introduction of the marked point process setup. Section 3 reviews models for the case of failure data with a single type of events, with emphasis on virtual age models and trend-renewal processes. Section 4 is devoted to a discussion of unobserved heterogeneity in repairable systems data. The model cube for heterogeneous trend-renewal processes is considered in particular. In Section 5 we consider various approaches to trend testing, both for data coming from single systems and from several similar systems. The possible extension of virtual age models to the marked process case is considered in Section 6. This section is based on some recent papers on the subject. Some concluding remarks are given in Section 7, in particular concerning topics not covered in the main text.

## 2. NOTATION AND BASIC DEFINITIONS

We consider a repairable system where time usually runs from $t = 0$ and where events occur at ordered times $T_1, T_2, \ldots$. Here time is not necessarily calendar time, but can in principle be any suitable measurement which is nondecreasing with calendar time, such as operation time, number of cycles, number of kilometers run, length of a crack and so forth. As already mentioned in the Introduction, we shall disregard time durations of repair and maintenance, so we assume that the system is always restarted immediately after failure or a maintenance action. Types of events (type of maintenance, type of failure, etc.) are, when applicable, recorded as $J_1, J_2, \ldots$ with $J_i \in \mathcal{J}$ for some mark space $\mathcal{J}$ which will depend on the current application. For simplicity we will here always assume that $\mathcal{J}$ is a finite set. The observable process $(T_1, J_1), (T_2, J_2), \ldots$ will be called the marked event process or occasionally the failure process. The interevent, or interfailure, times will be denoted $X_1, X_2, \ldots$. Here $X_i = T_i - T_{i-1}, i = 1, 2, \ldots$, where for convenience we define $T_0 \equiv 0$. Figure 1 illustrates the notation. We also make use of the counting process representation $N_j(t)$ equal to the number of events of type $j$ in $(0,t]$, which counts the number of events of type $j \in \mathcal{J}$, and $N(t) = \sum_{j \in \mathcal{J}} N_j(t)$, which counts the number of events irrespective of their types.

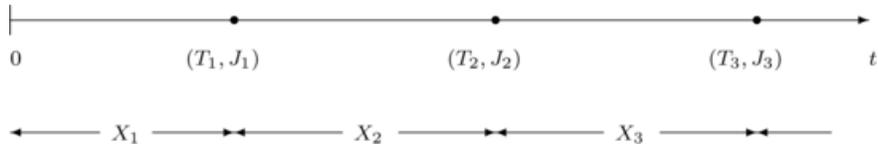

Fig. 1. *Event times $(T_i)$, event types $(J_i)$ and sojourn times $(X_i)$ of a maintained system.*

To describe probability models for repairable systems we use some notation from the theory of point processes. A key reference is Andersen, Borgan, Gill and Keiding [4]. Let $\mathcal{F}_{t-}$ denote the history of the marked event process up to, but not including, time $t$. In models without covariates we assume that $\mathcal{F}_{t-}$ includes all information on event times and event types before time $t$. Formally, $\mathcal{F}_{t-}$ is generated by the set $\{N_j(s) : 0 \leq s < t, j \in \mathcal{J}\}$.

Suppose then that a possibly time-dependent covariate vector $\mathbf{Z}(t)$ is observed for the system. In this case the covariate history $\{\mathbf{Z}(s) : 0 \leq s \leq t\}$ should be added to the history $\mathcal{F}_{t-}$ for each $t > 0$. This will imply that just before any time $t$ we have the complete information on the previous events, as well as the complete covariate history including the value of the covariate at time $t$. In the case of a time-constant covariate vector $\mathbf{Z}$, the information in $\mathbf{Z}$ is added to each history $\mathcal{F}_{t-}$.

The conditional intensity of the process with respect to events of type $j \in \mathcal{J}$ is now defined as

$$(1) \quad \gamma_j(t) = \lim_{\Delta t \downarrow 0} \frac{\Pr(\text{event of type } j \text{ in } [t, t+\Delta t)|\mathcal{F}_{t-})}{\Delta t},$$

which we call the type-specific intensity for $j$. Thus, $\gamma_j(t)\Delta t$ is approximately the probability of an event of type $j$ in the time interval $[t, t+\Delta t)$ given the history before time $t$. Further, we let $\gamma(t) = \sum_{j \in \mathcal{J}} \gamma_j(t)$ so that $\gamma(t)\Delta t$ is approximately the conditional probability of an event of any type in the time interval $[t, t+\Delta t)$, where it has been tacitly assumed that the probability of more than one event in an interval $[t, t+\Delta t)$ is $o(\Delta t)$. Note that the $\gamma_j(\cdot)$ and hence the $\gamma(\cdot)$ may be functions of the covariate vector $\mathbf{Z}(\cdot)$ when appropriate. In typical applications, $\gamma_j(t)$ may depend on the covariate history only through the value $\mathbf{Z}(t)$ at time $t$. Further, it is common to assume that $\gamma_j(t) = \gamma_j^0(t) g(\mathbf{Z}(t))$, with $\gamma_j^0(t)$ depending only on the pure event history $\{N_j(s) : 0 \leq s < t, j \in \mathcal{J}\}$, and with $g(\cdot)$ being some parametric function of the covariate vector such as the exponential one, $g(\mathbf{z}) = \exp(\boldsymbol{\beta}'\mathbf{z})$, where $\boldsymbol{\beta}$ is a parameter vector.

For statistical inference we need an expression for the likelihood function. Suppose that a single system with a marked event process as described above is observed from time 0 to time $\tau$, resulting in observations $(T_1, J_1), (T_2, J_2), \ldots, (T_{N(\tau)}, J_{N(\tau)})$, in addition to the covariate vector $\mathbf{Z}(s)$ for $0 \leq s \leq \tau$ if applicable. The likelihood function is then given by ([4], Section II.7)

$$(2) \quad L = \left\{\prod_{i=1}^{N(\tau)} \gamma_{J_i}(T_i)\right\} \exp\left\{-\int_0^\tau \gamma(u)\, du\right\}.$$

A rough verification of (2) can be given as follows. First, partition the interval $(0, \tau]$ into $s$ equal pieces, each of length $h = \tau/s$. Assume that $s$ is so large that at most one event can happen in an interval of length $h$. Then the conditional probability of an event of type $j$ in the interval $[(k-1)h, kh), k = 1, \ldots, s$, given the history before $(k-1)h$, is roughly $\gamma_j(kh)h$, while the conditional probability of no event in this interval is roughly $1 - \gamma(kh)h$. The probability of a realization of the process from 0 to $\tau$ will therefore include a product of $N(\tau)$ terms of the type $\gamma_j(kh)h$, corresponding to the observed events, and which in the limit as $h \to 0$ (after dividing by the normalization $h^{N(\tau)}$) gives the product term on the right-hand side of (2). The exponential part of (2) comes from taking the limit of the product of the terms $1 - \gamma(kh)h \approx \exp\{-\int_{(k-1)h}^{kh} \gamma(t)\, dt\}$ for the intervals that contain no event, assuming continuity of $\gamma(\cdot)$.

The likelihood function (2) is valid under the assumption that $\tau$ is a stopping time, which means that its value depends stochastically only on the past history. This property holds for the standard censoring schemes used in practice and in particular when $\tau$ is independent of the event process. There is, however, an increasing awareness of the need to allow for dependent censoring in many applications (see, e.g., [33]).

In typical applications, data will be available for several similar systems, with stopping times $\tau$ usually varying from system to system. Under the assumption of stochastic independence and identical probability mechanisms for the systems, the total likelihood will be the product of expressions (2) computed for all systems. For both parametric and nonparametric models of this kind there is a well developed theory for estimation based on the martingale approach to point processes ([4] gives a comprehensive account). Relevant references for statistical inference in reliability models are, among others, Ascher and Feingold [5], Rausand and Høyland [59], Crowder, Kimber, Smith and Sweeting [21] and Meeker and Escobar [52].

## 3. MODELS FOR REPAIRABLE SYSTEMS WITH A SINGLE TYPE OF EVENT

In the present section we assume that the observations are just the failure times $T_1, T_2, \ldots$. Thus the mark space $\mathcal{J}$ will be ignored.

A large number of models can be obtained in terms of a given hazard function $z(t)$, which we think of as being the hazard function of the time to first failure of a new system. The corresponding density and cumulative distribution function are denoted, respectively, $f(t)$ and $F(t)$, so $z(t) = f(t)/(1 - F(t))$. The idea is to use the function $z(t)$ together with a specification of the repair strategy to define the conditional intensity function $\gamma(t)$ of the failure process. Models of this type are considered in Sections 3.1 and 3.2. The corresponding models may be extended to the case with observed covariates, although this will not be made explicit. As described in Section 2, the conditional intensities of the form $\gamma(t)$ as considered below may be multiplied with a factor $g(\mathbf{Z}(t))$ that defines the dependence of the covariate value at time $t$.

### 3.1 Perfect and Minimal Repair Models

Suppose first that after each failure, the system is repaired to a condition as good as new. In this case the failure process is modeled by a renewal process with interevent time distribution $F$, denoted $\mathrm{RP}(F)$. Clearly

$$\gamma(t) = z(t - T_{N(t-)}),$$

where $t - T_{N(t-)}$ is the time since the last failure strictly before time $t$.

Suppose instead that after a failure, the system is repaired only to the state it had immediately before the failure, called a minimal repair. This means that the conditional intensity of the failure process immediately after the failure is the same as it was immediately before the failure, and hence is exactly as it would be if no failure had ever occurred. Thus we must have

$$\gamma(t) = z(t),$$

so that the process is a NHPP with intensity $z(t)$, denoted $\mathrm{NHPP}(z(\cdot))$. In practice a minimal repair usually corresponds to repairing or replacing only a minor part of the system.

### 3.2 Imperfect Repair Models and the Virtual Age of a System

A classical model, suggested by Brown and Proschan [13], assumes that at the time of each failure a perfect repair occurs with probability $p$ and a minimal repair occurs with probability $1 - p$, independently of the previous failure history. This model is a simple example of what has been called an imperfect repair model, and was later generalized in several directions.

Kijima [34] suggested two imperfect repair models, both involving what is called the virtual age (or effective age) of the system. The idea is to distinguish between the system's age, which is the time elapsed since the system was new, usually at time $t = 0$, and the virtual age of the system, which describes its present condition when compared to a new system. The virtual age is redefined at failures according to the type of repair performed and it runs along with the true time between repairs. More precisely, a system with virtual age $v \geq 0$ is assumed to behave exactly like a new system which has reached age $v$ without having failed. The hazard rate of a system with virtual age $v$ is thus $z_v(t) = z(v + t)$ for $t > 0$, where $z(\cdot)$ is the hazard rate of the time to first failure of the system.

It should be clear at this stage that models based on virtual ages make sense only if the underlying hazard functions $z(\cdot)$ are nonconstant. In fact, if $z(\cdot)$ is constant, then a reduction of virtual age would not influence the rate of failures.

A variety of imperfect repair models can be obtained by specifying properties of the virtual age process in addition to the hazard function $z(t)$ of a new system. For this, suppose $v(i)$ is the virtual age of the system immediately after the $i$th event, $i = 1, 2, \ldots$. The virtual age at time $t > 0$ is then defined by $A(t) = v(N(t-)) + t - T_{N(t-)}$, which is the sum of the virtual age after the last event before $t$ and the time elapsed since the last event. The process $A(t)$, called the virtual age process by Last and Szekli [40], thus increases linearly between events and may jump only at events. It follows that

$$(3) \quad \gamma(t) = z_{v(N(t-))}(t - T_{N(t-)}) = z(A(t)),$$

assuming that $A(t)$ is included in $\mathcal{F}_{t-}$ for all $t$. This means in turn that $v(i)$ is contained in $\mathcal{F}_{T_i}$ for each $t$ so that $v(i)$ depends on the history up to and including $T_i$. The likelihood then becomes

$$L = \left\{ \prod_{i=1}^{N(\tau)} z(v(i-1) + X_i) \right\}$$
$$\cdot \exp\left\{ -\sum_{i=1}^{N(\tau)} \int_0^{X_i} z(v(i-1) + u) \, du \right.$$
$$\left. - \int_0^{\tau - T_{N(\tau)}} z(v(N(\tau)) + u) \, du \right\}.$$

This can be recognized as being the same as

$$\left\{ \prod_{i=1}^{N(\tau)} f_{v(i-1)}(X_i) \right\} \{ 1 - F_{v(N(\tau))}(\tau - T_{N(\tau)}) \},$$

where $f_v(t) = f(v+t)/(1-F(v))$ and $F_v(t) = (F(v+t) - F(v))/(1 - F(v))$ are, respectively, the density and the cumulative distribution function of time to next failure for a system with virtual age $v$ and hence with hazard rate $z_v(\cdot)$.

It is clear that the perfect repair and minimal repair models are the special cases where, respectively, $v(i) = 0$ and $v(i) = T_i, i = 1, 2, \ldots$. In Kijima's [34] model I, the virtual age $v(i)$ equals $\sum_{k=1}^{i} D_k X_k$, where $D_1, D_2, \ldots$ is a sequence of random variables on the interval $[0, 1]$ such that $D_k$ is independent of $\mathcal{F}_{T_k-}$ for each $k$. Note that $\mathcal{F}_{T_k-}$ includes $D_1, D_2, \ldots, D_{k-1}$ so that in particular the $D_k$ are independent. In Kijima's model II the virtual age $v(i)$ is set to $\sum_{k=1}^{i} (\prod_{j=k}^{i} D_j) X_k$ with the same conditions for the $D_k$. This means that the virtual age after the $i$th failure equals $D_i$ multiplied by the virtual age of the system just prior to the $i$th failure. The model of Brown and Proschan [13] is obtained when $D_i$ is 1 with probability $1 - p$ and 0 with probability $p$ for all $i$.

Dorado, Hollander and Sethuraman [22] studied nonparametric statistical inference in a model slightly more general than Kijima's models described above. Nonparametric statistical inference in the Brown–Proschan model was first studied by Whitaker and Samaniego [63] and later by Hollander, Presnell and Sethuraman [31].

Recall that for the above models, the $D_i$ need to be observed for likelihood inference using (2) to be valid. This means in effect that the type of repair (minimal or perfect) must be reported for each repair action. In real applications, however, exact information on the type of repair is rarely available. The estimation problem in the case of unobserved $D_i$ has been considered by, for example, Lim [45] (suggesting an EM algorithm approach) and Langseth and Lindqvist [38, 39].

Doyen and Gaudoin [23] studied classes of virtual age models based on deterministic reduction of virtual age due to repairs, and hence not requiring the observation of repair characteristics. The basic models of this type can be obtained simply by letting the $D_i$ in Kijima's models above be replaced by parametric functions. A simple example of [23] is to use $1 - \rho$ for $D_i$, where $0 < \rho < 1$ is a so-called age reduction factor.

There is a large literature on reliability modeling using the virtual age process. For a review we refer to Pham and Wang [57] and Lindqvist [46]. Section 6 presents an attempt to define a multivariate virtual age process and corresponding repairable system models with several types of events.

### 3.3 Generalized Linear Model Types

Berman and Turner [10] considered estimation in parametric models with the conditional intensity being of the generalized linear model type

$$(4) \quad \gamma(t) = g\left\{ \sum_{i=0}^{p} \beta_i z_i(t) \right\},$$

where $g$ is a known monotonic continuous function, the $z_i(t)$ are known functions of $t$ and the history $\mathcal{F}_{t-}$, and the $\beta_i$ are unknown parameters. Note that the functions $z_i(t)$ may be functions of the covariates if available. One aim of the paper was to demonstrate how to use software for generalized linear models to analyze repairable systems data. The model

(4) is closely related to the modulated renewal process introduced in [18] for which Cox suggested a semiparametric approach for inference using a partial likelihood.

The special case of (4) obtained when $g(y) = e^y$ was applied to repairable systems by Lawless and Thiagarajah [43]. In particular, they considered the model

$$(5) \qquad \gamma(t) = e^{\beta_0 + \beta_1 g_1(t) + \beta_2 g_2(t - T_{N(t-)})},$$

where $g_1$ and $g_2$ are known functions. This conditional intensity is a function of both the calendar time and the time since last failure. Note that $\beta_1 = 0$ gives a RP and $\beta_2 = 0$ gives a NHPP, while $\beta_1 = \beta_2 = 0$ gives a HPP.

### 3.4 The Trend-Renewal Process

A class of processes called inhomogeneous gamma processes was suggested by Berman [9]. Berman motivated the inhomogeneous gamma process by first considering the process $T_1, T_2, \ldots$ obtained by observing every $\kappa$th event of a NHPP, where $\kappa$ is a positive integer. He then showed how to generalize to the case when $\kappa$ is any positive number.

We present now a generalization of Berman's idea, called the trend-renewal process, which was extensively studied by Lindqvist, Elvebakk and Heggland [48]. We will use this process in particular to describe the three dimensions related to the properties of repairable systems.

The idea behind the trend-renewal process is to generalize the following well-known property of the NHPP. First let the cumulative intensity function that corresponds to an intensity $\lambda(\cdot)$ be defined by $\Lambda(t) = \int_0^t \lambda(u)\,du$. Then if $T_1, T_2, \ldots$ is a NHPP$(\lambda(\cdot))$, the time-transformed stochastic process $\Lambda(T_1)$, $\Lambda(T_2), \ldots$ is HPP(1).

The trend-renewal process (TRP) is defined simply by allowing the above HPP(1) to be any renewal process RP$(F)$. Thus, in addition to the intensity function $\lambda(t)$, for a TRP we need to specify a distribution function $F$ of the interarrival times of this renewal process. Formally we can define the process TRP$(F, \lambda(\cdot))$ as follows:

Let $\lambda(t)$ be a nonnegative function defined for $t \geq 0$, and let $\Lambda(t) = \int_0^t \lambda(u)\,du$. The process $T_1, T_2, \ldots$ is called TRP$(F, \lambda(\cdot))$ if the transformed process $\Lambda(T_1), \Lambda(T_2), \ldots$ is RP$(F)$, that is, if the $\Lambda(T_i) - \Lambda(T_{i-1}), i = 1, 2, \ldots$, are i.i.d. with distribution function $F$. The function $\lambda(\cdot)$ is called the trend function, while $F$ is called the renewal distribution. To have uniqueness of the model, it is usually assumed that $F$ has expected value 1.

Figure 2 illustrates the definition. For a NHPP$(\lambda(\cdot))$, the RP$(F)$ would be HPP(1). Thus TRP$(1 - e^{-x}, \lambda(\cdot)) = $ NHPP$(\lambda(\cdot))$. Also, TRP$(F, 1) = $ RP$(F)$, which shows that the TRP class includes both the RP and NHPP classes.

As a motivation for the TRP model, suppose that failures of a particular system correspond to replacement of a major part, for example, the engine of a tractor (as in the data given by Barlow and Davis [6]), while the rest of the system is not maintained. Then if the rest of the system is not subjected to wear, a renewal process would be a plausible model for the observed failure process. In the presence of wear, on the other hand, an increased replacement frequency is to be expected. This is achieved in a TRP model by accelerating the internal time of the renewal process according to a time transformation $\Lambda(t) = \int_0^t \lambda(u)\,du$ which represents the cumulative wear. The TRP model thus has some similarities to accelerated failure time models.

It can be shown [48] that the conditional intensity function for the TRP$(F, \lambda(\cdot))$ is

$$(6) \qquad \gamma(t) = z(\Lambda(t) - \Lambda(T_{N(t-)}))\lambda(t),$$

where $z(\cdot)$ is the hazard rate that corresponds to $F$. This is a product of one factor, $\lambda(t)$, which depends on the age $t$ of the system and one factor which depends on a transformed time from the last previous failure. However, time since last failure is measured on a scale that depends on the cumulative intensity of failures. This shows that the TRP class does not contain, nor is contained in, the classes of processes considered in the previous subsection.

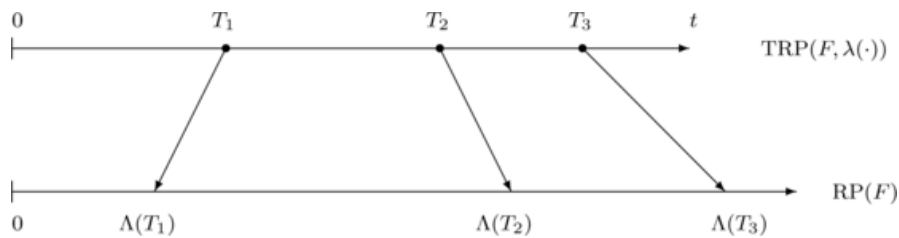

Fig. 2. *The defining property of the trend-renewal process.*

Suppose now that a single system has been observed in $[0, \tau]$, with failures at $T_1, T_2, \ldots, T_{N(\tau)}$. If a $\text{TRP}(F, \lambda(\cdot))$ is used as a model, then substitution of (6) into (2) gives the likelihood

$$
\begin{aligned}
L = {} & \left\{ \prod_{i=1}^{N(\tau)} z[\Lambda(T_i) - \Lambda(T_{i-1})]\lambda(T_i) \right\} \\
& \cdot \exp\left\{ -\int_0^\tau z[\Lambda(u) - \Lambda(T_{N(u-)})]\lambda(u)\,du \right\}.
\end{aligned}
\tag{7}
$$

Equivalently, if $f$ is the density function that corresponds to $F$, we can write this likelihood as

$$
\begin{aligned}
L = {} & \left\{ \prod_{i=1}^{N(\tau)} f[\Lambda(T_i) - \Lambda(T_{i-1})]\lambda(T_i) \right\} \\
& \cdot \{1 - F[\Lambda(\tau) - \Lambda(T_{N(\tau)})]\}.
\end{aligned}
\tag{8}
$$

The latter form of the likelihood follows directly from the definition of the TRP, since the conditional density of $T_i$ given $T_1 = t_1, \ldots, T_{i-1} = t_{i-1}$ is $f[\Lambda(t_i) - \Lambda(t_{i-1})]\lambda(t_i)$, and the probability of no failures in the time interval $(T_{N(\tau)}, \tau]$, given $T_1, \ldots, T_{N(\tau)}$, is $1 - F[\Lambda(\tau) - \Lambda(T_{N(\tau)})]$.

A possible extension of the TRP to include covariates would be to multiply the trend function $\lambda(t)$ by a factor $g(\mathbf{Z}(t))$, for example, of the form $\exp(\boldsymbol{\beta}'\mathbf{Z}(t))$ as suggested in Section 2. The $\lambda(t)$ would then play the role of a baseline trend function. This definition generalizes in a natural way the commonly used NHPP model with covariates; see, for example, [41].

## 4. UNOBSERVED HETEROGENEITY IN REPAIRABLE SYSTEMS

Analyses of reliability data often lead to an apparent decreasing failure rate which could be counterintuitive in view of wear and aging effects. Proschan [58] pointed out that such observed decreasing rates could be caused by unobserved heterogeneity. Proschan presented failure data from 17 air conditioner systems on Boeing 720 airplanes. Applying Mann's [51] nonparametric trend test to each system and then combining to a global test statistic, he argued that there is no significant trend in the failure times for each separate plane. He then concluded by a similar test that "it seems safe to accept the exponential distribution as describing the failure interval, although to each plane may correspond a different failure rate." He demonstrated this last fact statistically by using a result from Barlow, Marshall and Proschan [7] which implies that a mixture of exponential distributions has a decreasing failure rate. More precisely, he applied again the Mann test, which is sensitive to a decreasing failure rate, on the pooled interfailure times from all the planes. In this way he obtained a $p$-value of 0.007 for the null hypothesis of identical exponential distributions of the interfailure times.

Heterogeneity in connection with Poisson processes was in fact studied as early as 1920 by Greenwood and Yule [27], who used a compound Poisson distribution. Later, Maguire, Pearson and Wynn [50], studying occurrences of industrial accidents, showed how Laplace transforms enter general expressions for resulting distributions of intervals and counts. Cox [17] considered the possibility of heterogeneity, which he called variance components, between homogeneous Poisson processes and listed several reasons for the interest in such models for repairable systems data.

It has similarly long been known in biostatistics that neglecting individual heterogeneity may lead to severe bias in estimates of lifetime distributions. The idea is that individuals or components have different "frailties" and that those who are most "frail" will die or fail earlier than the others. This in turn leads to a decreasing population hazard, which has often been misinterpreted. Important references on heterogeneity in the biostatistics literature are [62], [32] and [2]. It should be noted that heterogeneity is, in general, unidentifiable if it is considered as an individual quantity. For identifiability it is necessary that frailty be common to several individuals, for example, in family studies in biostatistics, or if several events are observed for each individual, such as for the repairable systems considered in this paper and more generally for recurrent events data. The presence of heterogeneity is often apparent for data from repairable systems if there is a large variation in the number of events per system. However, it is not really possible to distinguish between heterogeneity and dependence of the intensity on past events for a single process. It is a fact, though, that ignorance of an existing heterogeneity may lead to suboptimal or even wrong decisions.

### 4.1 Modeling Heterogeneity for Repairable Systems

The common way to model heterogeneity is to include an unobservable multiplicative constant in the conditional intensity of the process; see, for example, [62]. For systems with a single type of event this is done by first replacing the conditional intensities $\gamma(t)$ in (1) by $a\gamma(t)$, where $a$ is a random variable that represents the frailty of the system and such that $a$ is included in $\mathcal{F}_{t-}$ for each $t$. Note that $\gamma(t)$ as described in Section 2 may well be a function of covariates. Now $a$ can be viewed as being the effect of an unobserved covariate. Systems with a large value of $a$ will have a larger failure proneness than systems with a low value of $a$. Intuitively, the variation in the $a$ between systems implies that the variation in observed number of failures among the systems is larger than would be expected if the failure processes were identically distributed. Now, since $a$ is unobservable, one needs to take the expectation of the likelihood that results from (2) with respect to the distribution of $a$ in order to have a likelihood function for the observed data.

In the marked point process formulation of Section 2 we may more generally assume that there are different frailty variables for each event type $j \in \mathcal{J}$. More precisely, we assume that there is a random vector $\mathbf{a} = (a_j, j \in \mathcal{J})$ such that the type-specific intensities for given $\mathbf{a}$ are $a_j \gamma_j(t)$, respectively, where $\gamma_j(t)$ corresponds to the type-specific conditional intensity defined in Section 2. The resulting likelihood including heterogeneity is thus

$$L = E_{\mathbf{a}}\left[\left(\prod_{i=1}^{N(\tau)} a_{J_i} \gamma_{J_i}(T_i)\right) \cdot \exp\left\{-\sum_{j \in \mathcal{J}} a_j \int_0^\tau \gamma_j(u)\,du\right\}\right], \quad (9)$$

where the expected value is taken with respect to the joint distribution of $\mathbf{a}$. Multivariate frailty distributions are considered by, for example, Hougaard [32] and Aalen [1].

In the case of several independent systems, it is assumed that the $\mathbf{a}$'s that correspond to the systems are i.i.d. from the given joint distribution. The total likelihood is then the product of factors (9), one for each system. Note that for identifiability it may be necessary to introduce a normalization of $\mathbf{a}$, for example, to assume that $E(\|\mathbf{a}\|) = 1$. This is because otherwise a scale factor may be moved from $a_j$ to $\gamma_j(\cdot)$ or vice versa without changing the value of (9). Alternatively one may let the $a_j$ act as free random scale parameters in the model if the $\gamma_j(\cdot)$ themselves do not include scale parameters.

For the special case of a single type of event one obtains simplification of the likelihood function in (9),

$$L = E_a\left[a^{N(\tau)} \left(\prod_{i=1}^{N(\tau)} \gamma(T_i)\right) \cdot \exp\left\{-a \int_0^\tau \gamma(u)\,du\right\}\right], \quad (10)$$

where the expectation is with respect to the distribution of the random variable $a$ and where for normalization one will usually assume $E(a) = 1$.

Expression (10) suggests that a gamma distribution for $a$ is mathematically convenient, since a closed form expression of the likelihood is obtained. More generally, for the version (9), a multivariate gamma distribution for $\mathbf{a}$ leads to a simplified expression (see, e.g., [1] and [32] regarding multivariate gamma distributions).

Consider now the likelihood (10) and suppose that $a$ is gamma distributed with $E(a) = 1, \text{Var}(a) = \delta$. Then a straightforward computation gives

$$\begin{aligned}L &= \left\{\prod_{i=1}^{N(\tau)} \gamma(T_i)\right\} \\ &\quad \cdot \frac{\Gamma(N(\tau) + 1/\delta)}{\delta^{1/\delta}\Gamma(1/\delta)[1/\delta + \int_0^\tau \gamma(u)\,du]^{N(\tau)+1/\delta}} \\ &= \left\{\prod_{i=1}^{N(\tau)} \gamma(T_i)\right\} \\ &\quad \cdot \frac{[\delta(N(\tau)-1)+1][\delta(N(\tau)-2)+1]\cdots 1}{[\delta \int_0^\tau \gamma(u)\,du + 1]^{N(\tau)+1/\delta}},\end{aligned} \quad (11)$$

where we have used the fact that $\Gamma(r+1) = r\Gamma(r)$. Recall that $\gamma(T_i)$ may well include covariates. This likelihood expression is applicable, for example, together with the virtual age model (3) and the generalized linear model types (4) and (5). It is also the likelihood function for NHPPs with heterogeneity and possibly covariates, as studied in Lawless [41], and results in the likelihood of the so-called compound power law model studied by Engelhardt and Bain [26].

We remark that (11) converges to (2) (assuming a single type of event) as $\delta \to 0$.

### 4.2 Heterogeneity in the TRP Model, the HTRP Model

Lindqvist, Elvebakk and Heggland [48] introduced heterogeneity into the TRP model by including an unobservable random multiplicative constant $a$ in the trend function $\lambda(t)$, thus considering the conditional model $\mathrm{TRP}(F, a\lambda(\cdot))$ with a renewal distribution $F$ that does not depend on $a$. This definition is consistent with the regression version of TRP as suggested at the end of Section 3.4. Now the $a$ replaces the function $g(\mathbf{Z}(t))$ used there. Note that in practice one may want to include both the frailty $a$ and a covariate factor $g(\mathbf{Z}(t))$. To simplify the discussion, we will, however, not consider covariates in our presentation.

Considering (6), it is seen that the conditional intensity function given $a$ is no longer of the simple multiplicative form $a\gamma(t)$ which was assumed in the previous subsection. This is because the $\Lambda(\cdot)$ in (6) is also multiplied by $a$. Instead of the expression (10), the relevant likelihood from one system becomes, using (7),

$$L = E_a\left[\left\{\prod_{i=1}^{N(\tau)} z[a(\Lambda(T_i) - \Lambda(T_{i-1}))]a\lambda(T_i)\right\}\right.$$
(12) $$\left.\cdot \exp\left\{-a\int_0^\tau z[a(\Lambda(u) - \Lambda(T_{N(u-)}))]\right.\right.$$
$$\left.\left.\cdot \lambda(u)\,du\right\}\right]$$

or, using (8),

$$L = E_a\left\{\prod_{i=1}^{N(\tau)} f[a(\Lambda(T_i) - \Lambda(T_{i-1}))]a\lambda(T_i)\right\}$$
(13)
$$\cdot \{1 - F[a(\Lambda(\tau) - \Lambda(T_{N(\tau)}))]\}.$$

Here $f$ and $z$ are, respectively, as before, the density and hazard function of the distribution $F$.

The expressions (12) and (13) appear to be less tractable than the expression (10). Lindqvist, Elvebakk and Heggland [48] obtained, however, a rather simple expression for the likelihood in the case of an inhomogeneous gamma process with gamma distributed heterogeneity factor $a$, under the further assumption that the stopping times $\tau$ coincide with failure times. In this case the last factor of (13) disappears, and letting $F$ be the gamma distribution with unit expectation and variance $\gamma$, while $a$ is gamma distributed with unit expectation and variance $\delta$, one obtains

$$L = \left\{\prod_{i=1}^{N(\tau)} (\Lambda(T_i) - \Lambda(T_{i-1}))^{1/\gamma - 1}\lambda(T_i)\right\}$$
$$\cdot (\Gamma(N(\tau)/\gamma + 1/\delta))$$
$$\cdot \left\{\gamma^{N(\tau)/\gamma}[\Gamma(1/\gamma)]^{N(\tau)}\delta^{1/\delta}\right.$$
$$\left.\cdot \Gamma(1/\delta)[1/\delta + (1/\gamma)\Lambda(T_{N(\tau)})]^{N(\tau)/\gamma + 1/\delta}\right\}^{-1}.$$

Note that for $\gamma = 1$ this is of the same form as in (11).

We use the notation $\mathrm{HTRP}(F, \lambda(\cdot), H)$ for the model with likelihood (12) or, equivalently, (13). The "H" in HTRP here stands for heterogeneity, and the $H$ which is added to $(F, \lambda(\cdot))$ in the notation is the distribution of the variable $a$, which can be any positive distribution with expected value 1.

### 4.3 The Three Dimensions of a Repairable System Description: The Model Cube and the Log-Likelihood Cube

A useful feature of the HTRP model is that several models for repairable systems can be represented as submodels. With the notation HPP, NHPP, RP and TRP used as before, and with an H in front meaning the model which includes heterogeneity, Figure 3 shows how the HTRP and the seven sub-models can be represented in a cube [25]. Each vertex of the cube represents a model, and the lines that connect them correspond to changing one of the three "coordinates" $(F, \lambda(\cdot), H)$ in the HTRP notation. Going to the right corresponds to introducing a time trend, going upward corresponds to entering a non-Poisson (renewal) case and going backward (inward) corresponds to introducing heterogeneity.

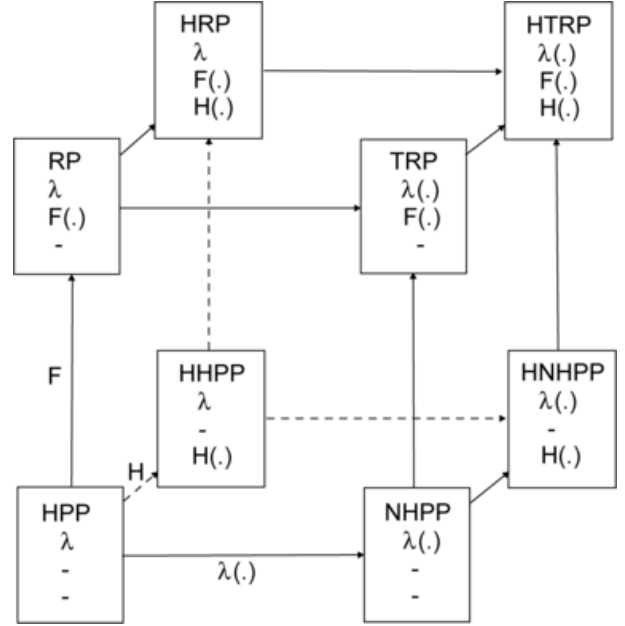

FIG. 3. *The model cube illustrating the* $\mathrm{HTRP}(F, \lambda(\cdot), H)$ *and the submodels obtained by restricting one or more of* $F, \lambda(\cdot), H$ *to their basic versions, respectively, $F$ being standard exponential (using the notation - in the figure), $\lambda(t) \equiv \lambda$ being constant in time and $H$ being the distribution deterministic at 1 (- in the figure).*

In analyzing data by parametric HTRP models we may use the cube to facilitate the presentation of maximum log-likelihood values and parameter estimates for the different models in a convenient, visual manner which may guide model choice (see [48]). Figures 4 and 5 show maximum likelihood values computed from the data of Proschan [58] and Aalen and Husebye [3], respectively. The latter data set is taken from a medical study and is included here to demonstrate results for data which are clearly non-Poisson distributed.

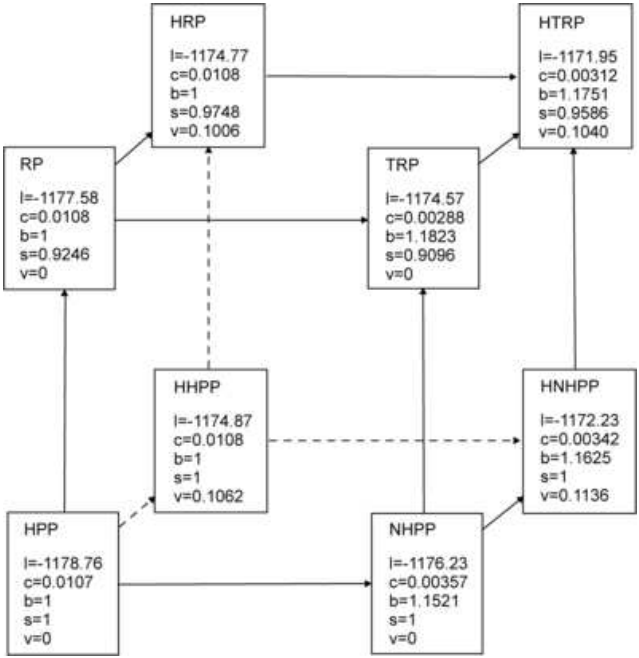

FIG. 4. *The log-likelihood cube for the data of Proschan* [58] *concerning failures of air conditioner systems on airplanes, fitted with a parametric* $\mathrm{HTRP}(F, \lambda(\cdot), H)$ *model and its submodels. Here $F$ is a Weibull distribution with expected value 1 and shape parameter $s$, $\lambda(t) = cbt^{b-1}$ is a power function of $t$ and $H$ is a gamma distribution with expected value 1 and variance $v$. The maximum value of the log-likelihood is denoted $l$.*

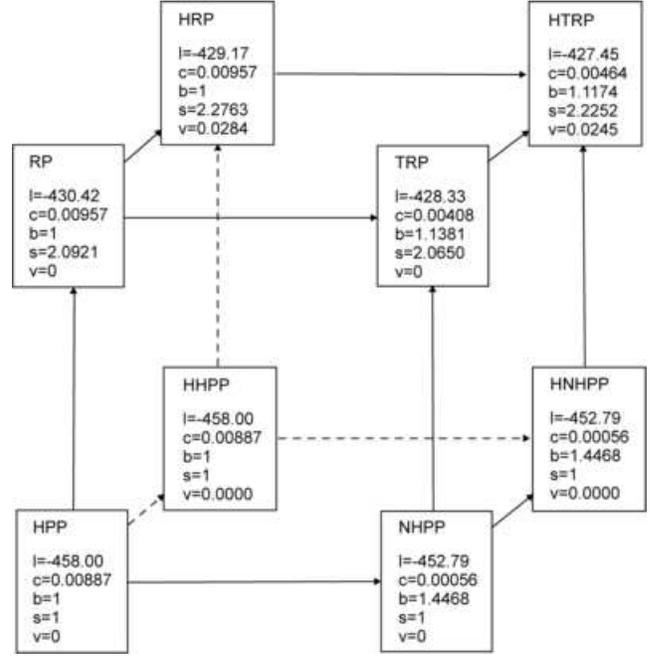

FIG. 5. *The log-likelihood cube for the data of Aalen and Husebye* [3] *concerning migratory motor complex periods, fitted with a parametric* $\mathrm{HTRP}(F, \lambda(\cdot), H)$ *model and its submodels. Here $F$ is a Weibull distribution with expected value 1 and shape parameter $s$, $\lambda(t) = cbt^{b-1}$ is a power function of $t$ and $H$ is a gamma distribution with expected value 1 and variance $v$. The maximum value of the log-likelihood is denoted $l$.*

For the Proschan data we conclude that the renewal distribution can be taken to be exponential, leaving us with the bottom face of the cube. Further, when comparing the front face to the back face there is clear reason to conclude that there is heterogeneity between the systems, with $\mathrm{Var}(a)$ being estimated to approximately 0.11. The conclusions so far are thus in accordance with the conclusions of Proschan [58]. However, a comparison of the left and right faces of the cube reveals a slight time trend. In fact, twice the log-likelihood difference from HHPP to HNHPP amounts to 5.28, giving a $p$-value of 0.022 assuming a chi-squared distribution with one degree of freedom of the corresponding likelihood ratio test statistic. The power parameter $b$ of the trend function is, furthermore, estimated as 1.16.

The most obvious conclusion for the Aalen and Husebye [3] data is that the renewal distribution is not exponential, implying that the upper face of the cube applies. Further, the differences in log-likelihood obtained by introducing heterogeneity are seen to be small enough to conclude there is no significant heterogeneity. However, as for the Proschan data, there seems to be a slight time trend. Here, twice the log-likelihood difference from RP to TRP amounts to 4.18, giving a $p$-value of 0.041, while the power parameter $b$ is estimated as 1.14 for the TRP model. Note the large difference in log-likelihood value between, for example, the TRP and NHPP models. As shown by the parameter estimates (Figure 5), the NHPP estimates seem to compensate for the large estimated shape parameter for the renewal distribution of the TRP by increasing the power parameter $b$ of the trend function (from 1.14 to 1.45). It is also seen that for the Poisson models (bottom face) there is no gain in log-likelihood by introducing heterogeneity. Thus the maximum likelihood estimates of the heterogeneity variance $v$ are given by the border value 0. This is so since the profile likelihood of $v$ can be shown to be a decreasing function of $v > 0$ near 0 (see [48] for a further discussion of this effect).

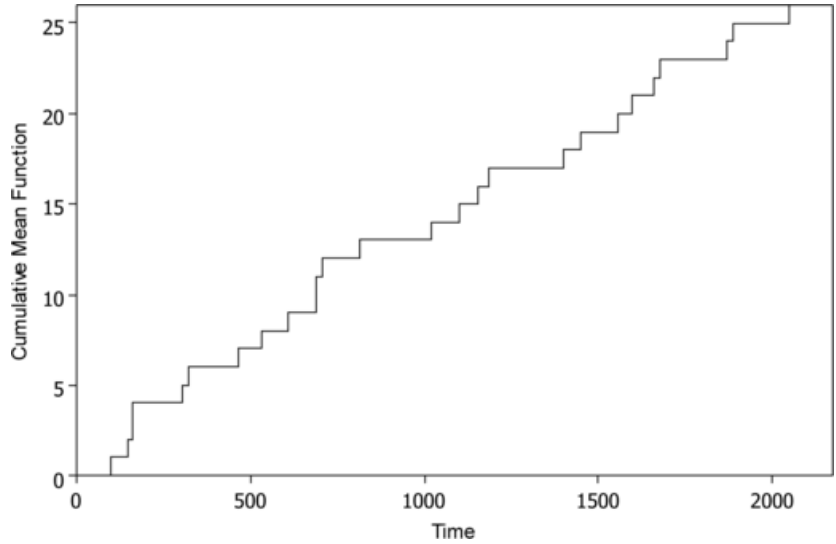

Fig. 6. *Plot of cumulative number of failures, $N(t)$, for air conditioner failures of plane* 7913 *in the Proschan* [58] *data.*

## 5. TREND TESTING

In many applications involving repairable systems, the main aim is to detect trends in the pattern of failures that occur over time. These may often be revealed as monotonic trends in the interfailure times, corresponding to either improving or deteriorating systems. Various types of nonmonotonic trends may also be present, for example, a cyclic trend or a bathtub shaped trend.

### 5.1 Graphical methods

A simple but informative way to check for a possible trend in the pattern of failures is to study plots like Figure 6, which is a plot of cumulative failure number versus failure time for a single system. The underlying data are failures of the air conditioner system of airplane 7913 of the Proschan [58] data. A convex plot would be indicative of a deteriorating system, while a concave plot would indicate an improving system. In Figure 6 there seems to be no significant deviation from a straight line, however, thus indicating no trend in interfailure times.

5.1.1 *Nelson–Aalen plot.* The plot of Figure 6 is a special case of the Nelson–Aalen plot to be described next. Assume that $m$ systems are observed, with the individual failure processes being independent and identically distributed. Suppose further that the $i$th process is observed on the time interval $(0, \tau_i]$ and let $y(t)$ denote the number of processes under observation at time $t$. Note that $y(t)$ is a function of the $\tau_i$ and not of the failure times. Let $T_k$ denote the $k$th arrival time in the superposed process, that is, $T_i$ is a failure time in one of the processes and $0 < T_1 \leq T_2 \leq \cdots \leq T_N \leq \tau$, where $\tau = \max\{\tau_i : i = 1, \ldots, m\}$. Define the cumulative mean function of a single process to be $M(t) = E(N(t))$. The Nelson–Aalen estimator of $M(t)$ is given by

$$\hat{M}(t) = \sum_{T_k \leq t} \frac{1}{y(T_k)},$$

where the sum is taken over all failure times $T_k$ before or at time $t$. Figure 7 shows the plot of $\hat{M}(t)$ for the data on times of valve-seat replacements in a fleet of $m = 41$ diesel engines, taken from [53]. The plot indicates that the replacement frequency is fairly constant up to 550 days and then increases as revealed from the convex shape of the curve at the right end.

The plot as defined here is studied, for example, in [53] and [42]. These papers also derive robust nonparametric estimates of the variance of $\hat{M}(t)$, valid under any distributional properties of the individual processes $N(t)$.

5.1.2 *TTT plot.* Consider the special case of the above where the $m$ processes are independent NHPPs with a common intensity function $\lambda(t)$. The superposed process is now a NHPP with intensity function $\phi(t) = \lambda(t) y(t)$, and hence (see Section 3.4) the process $\int_0^{T_1} \phi(u)\,du, \int_0^{T_2} \phi(u)\,du,\ldots$ is HPP(1) on $(0, \tau)$. Define the total time on test (TTT) at time $t$ by

$$r(t) = \int_0^t y(u)\,du.$$

Barlow and Davis [6] introduced the TTT plot for repairable systems data as a plot of the points

$$\left(\frac{i}{N}, \frac{r(T_i)}{r(\tau)}\right), \quad i = 1,\ldots,N.$$

The idea is that if $\lambda(t)$ is a constant, so that the processes are HPP, then the $r(T_i)/r(\tau), i = 1,\ldots,N$, form a HPP(1) on $[0,1]$. In this case the TTT plot is by its definition expected to be located near the main diagonal of the unit square. Under the alternatives of decreasing, increasing and bathtub-shaped intensity $\lambda(t)$, on the other hand, the TTT plots appear to be, respectively, convex, concave and S-shaped. Figure 8 shows the TTT plot of the valve-seat replacement data of Nelson [53]. The plot appears to be fairly straight, but with a slightly concave shape near the end corresponding to the increasing intensity here as revealed by the Nelson–Aalen plot in Figure 7.

### 5.2 Statistical Trend Tests

Statistical trend tests for repairable systems data were extensively discussed by Ascher and Feingold [5], Chapter 5B. A trend test is a statistical test for the null hypothesis that the failure process is stationary, in some sense to be made precise, versus alternatives that depend on the kind of trend one would like to detect. Here we give main attention to the null hypothesis that the process is a HPP or more generally a RP. However, as will be discussed below, some care should be taken when determining the relevant null hypothesis.

The null hypothesis of HPP is the most common and often the most useful in reliability applications. The corresponding null property, under the name "randomness," was studied in several papers in the 1950s, and various tests for randomness in time were devised. Here randomness pertained to the property that counts in given time intervals are Poisson-distributed. Maguire, Pearson and Wynn [50], however, discussed the advantages of using interevent times rather than counts to test for changes with time of the occurrence rate of events. Cox [17] stated eight different kinds of possible alternatives to randomness, one of them being trend in the sense that the conditional intensity is a smooth function of time.

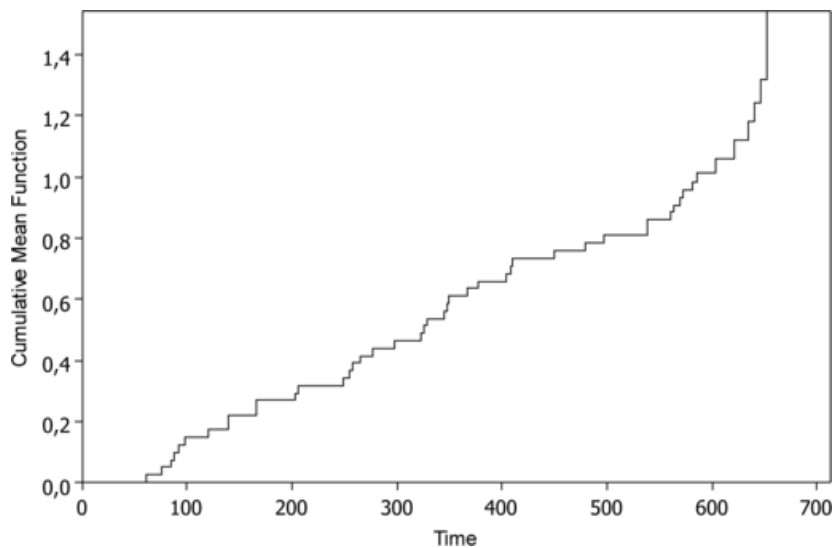

FIG. 7. *Nelson–Aalen plot of the estimated cumulative mean function $\hat{M}(t)$ for the valve-seat replacement data as given by Nelson* [53].

5.2.1 *Tests of the null hypothesis of HPP. Single process.* Suppose first that the null hypothesis is "the process is a HPP," with the alternative being a NHPP with monotone intensity. Two classical trend tests for this case are the Laplace test and the Military Handbook test (see, e.g., [5], page 79). To see how they are obtained, consider a single system observed on $[0, \tau]$. If the failure process is a HPP, then given $N(\tau) = n$, the failure times $T_1, T_2, \ldots, T_n$ are distributed as the ordering of $n$ i.i.d. uniform random variables on $[0, \tau]$. Equivalently, the $T_i/\tau$ ($i = 1, \ldots, n$) are distributed as ordered i.i.d. uniforms on $[0, 1]$ conditionally given $N(\tau) = n$. From this we can in principle obtain trend tests from any test for detecting deviations from a uniform sample. The Laplace test statistic is simply a normalization of $\sum_{i=1}^{n} T_i$, while the Military Handbook test statistic is similarly a normalization of $\sum_{i=1}^{n} \log T_i$. The Laplace test and the Military Handbook test are optimal tests against the alternatives of NHPPs with, respectively, log linear intensity and power intensity functions ([5], page 79).

*Several processes.* As in Section 5.1.2, assume that $m$ independent NHPPs with a common intensity function $\lambda(t)$ are observed, where the $i$th process is observed on the time interval $(0, \tau_i]$. Recall that, under the null hypothesis that $\lambda(t)$ is a constant, the $r(T_i)/r(\tau), i = 1, \ldots, N$, form a HPP(1) on $[0, 1]$. Kvaløy and Lindqvist [35] suggested from this that formal trend tests could be defined by substituting the $r(T_i)/r(\tau)$ into the Laplace and Military Handbook test statistics. While these TTT-based tests are powerful against monotone alternatives, the authors suggested using a test statistic based on the Anderson–Darling statistic as a general test with power against several kinds of trend.

For many applications, the null hypothesis needs to be weakened to state that each process is a HPP, but that intensities may differ from system to system. For example, in the data of Proschan [58], one may be interested in a simultaneous trend test for the systems, allowing there to be heterogeneities between them. Kvaløy and Lindqvist [35] suggested tests for this case called combined tests. A precise setting for these tests was recently defined by Kvist, Andersen and Kessing [37], who considered a model where the conditional intensity function for a particular system is given by $ag(\mathbf{Z})\lambda(t)$, where $a$ is an unobservable frailty variable as considered in Section 4.1, $\mathbf{Z}$ is a fixed-time covariate vector observed for each system, $g$ is a parametric regression function and $\lambda(t)$ is a baseline intensity function. Suppose that such a process is observed on the time interval $[0, \tau]$ with events at times $T_1, T_2, \ldots, T_{N(\tau)}$. Then, conditional on $(a, \mathbf{Z}, \tau, N(\tau))$, the $T_1/\tau, T_2/\tau, \ldots,$ $T_{N(\tau)}/\tau$ are distributed as $N(\tau)$ ordered standard uniform variables on $[0,1]$ if $\lambda(t)$ is constant. We are hence back to the setting of the beginning of this subsection. In practice one observes $m$ independent processes of this kind, with a common $\lambda(t)$, with the $a$ being i.i.d. unobservable random variables and the $\mathbf{Z}$ being observed covariate vectors for each system. The above-mentioned combined tests by Kvaløy and Lindqvist [35] can thus be used to test the null hypothesis that $\lambda(t)$ does not depend on $t$. Kvist, Andersen and Kessing [37] applied the Laplace type test of this kind on data from the Danish register on psychiatric hospital admissions.

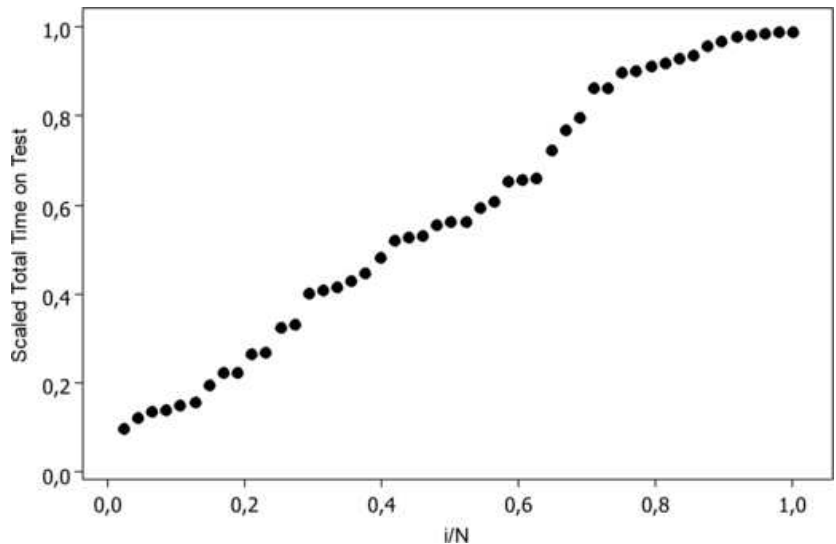

FIG. 8. *TTT plot of valve-seat data as given in [53].*

5.2.2 *Tests of the null hypothesis of RP.* The Laplace test and the Military Handbook test are tests for the null hypothesis that the data come from HPPs. Thus rejection of the null hypothesis means merely that the process is not a HPP. It could still, however, be a RP and thus still have "no trend." Lawless and Thiagarajah [43] and Elvebakk [25] concluded from simulations that the Laplace and Military Handbook tests in fact may be seriously misleading when used to detect trend departures from general renewal processes. Similarly, Lewis and Robinson [44] noted that these tests are not able to discriminate properly between trends in the data and the appearance of sequences of very long intervals.

To test the null hypothesis of RP, Lewis and Robinson [44] suggested modifying the Laplace test by dividing the test statistic by an estimate of the coefficient of variation of the interfailure times under the null hypothesis of a RP. This test, called the Lewis–Robinson test, is thus a simple modification of the Laplace test. Another classical trend test for the null hypothesis of RP is the rank test developed by Mann [51] and used by Proschan [58] (see Section 4).

Kvaløy and Lindqvist [36] presented a general class of tests for renewal process versus both monotonic and nonmonotonic trend for which the Lewis–Robinson and a useful Anderson–Darling type test are special cases.

Elvebakk [25] demonstrated how tests for the null hypothesis of RP can be obtained from tests for the Poisson case by adjusting their critical values by resampling failure data under the RP hypothesis. The general conclusion of Elvebakk [25] was to recommend the use of such resampled trend tests whenever it is not clear that the failure processes are of Poisson type. In particular he showed in a simulation study that the resampled tests are usually favorable to the Lewis–Robinson test, and that they do not lose much power under NHPP alternatives when compared to the standard tests.

5.2.3 *Tests of the null hypothesis of stationary interfailure times.* Lewis and Robinson [44] presented a test for distinguishing between a general stationary sequence of interfailure times $X_i$ and a monotonic trend in interfailure times. Elvebakk [25] extended the resampling trend testing approach described in the previous subsection, to cover the case when "no trend" corresponds to stationary interfailure times. The idea is to resample data under this new null hypothesis assumption. Elvebakk did this both by a parametric approach assuming an underlying autoregressive model and by employing a block bootstrap technique adapted from Hall [28]. Simulations indicated rather satisfactory performance of the method.

5.2.4 *Trend tests obtained as likelihood ratio tests.* In parametric models which include separate parameters for trend, trend tests may be performed as likelihood ratio tests that involve these parameters. An example is to test the null hypothesis $\beta_1 = 0$ in (5) which was suggested in [43]. Trend tests can also be obtained in models of the form $\text{HTRP}(F, \lambda(\cdot), H)$ by testing the null hypothesis that $\lambda(\cdot) \equiv \lambda$ using likelihood ratio tests. Note that this leads to tests of the null hypothesis that the processes are all renewal processes with a possibility of heterogeneity.

A nonparametric likelihood ratio test for the null hypothesis of a HPP versus the alternative of a NHPP with monotone intensity $\lambda(\cdot)$ was derived by Boswell [11]. A generalization to the null hypothesis of RP can be obtained using the nonparametric monotone estimator of $\lambda(\cdot)$ in the TRP model derived by Heggland and Lindqvist [29].

## 6. REPAIRABLE SYSTEMS WITH SEVERAL TYPES OF EVENTS

In this section we consider the general marked event process described in Section 2. The purpose is to show how new classes of maintenance and repair models can be obtained by generalizing the approach of the imperfect repair models for single type events considered in Section 3.2. To simplify the presentation we shall not allow covariates or heterogeneity in the models considered here.

As in Section 3.2, we consider first a nonrepairable unit. Assume that this unit may fail due to one of several causes or may be stopped for PM before it fails, in which case failure is prevented.

We can formally think of this as having a system with, say, $n$ components, denoted $\{C_1, C_2, \ldots, C_n\}$, where a unique failing component can be identified at failures of the system and where PM, if applicable, is represented by one of these components so as to simplify notation. Let $W_j$ be the potential failure time due to failure of component $C_j$, $j = 1, 2, \ldots, n$. What is observed is the failure time $T = \min(W_1, \ldots, W_n)$ and the identity of the failing component, say $J = j$ if the component $C_j$ fails. This determines a competing risks situation with $n$ competing risks and with the observed outcome $(T, J)$ ([20], Chapter 3). The joint distribution of $(T, J)$ is thus identifiable from data, as are the so-called type-specific hazards defined by

$$(14) \quad h_j(t) = \lim_{\Delta t \downarrow 0} \frac{\Pr(t < T \leq t + \Delta t, J = j | T > t)}{\Delta t}.$$

However, neither the joint nor the marginal distributions of the individual potential failure times $W_1, \ldots, W_n$ are identifiable in general from observation of $(T, J)$ only. This follows from the so-called Cox–Tsiatis impasse; see [20], Chapter 7. On the other hand, these marginal and joint distributions are indeed of interest in reliability applications, for example, in connection with maintenance optimization. An example is given in the next paragraph.

Consider the setup of Cooke [15, 16] that involves a competing risks situation with a potential failure of a unit at some time $W_1$ and a potential action of preventive maintenance to be performed at time $W_2$. Thus $n = 2$, while $C_1$ corresponds to failure of the unit $(J = 1)$ and $C_2$ $(J = 2)$ corresponds to the action of PM. Knowing the marginal distribution of $W_1$ would be particularly important since it is the basic failure time distribution of the unit when there is no PM. However, as noted above, the marginal distributions of $W_1$ and $W_2$ are not identifiable unless specific assumptions are made on the dependence between $W_1$ and $W_2$. The most common assumption of this kind is that $W_1$ and $W_2$ are independent, in which case identifiability follows ([61]; [20], Chapter 7). However, this assumption is unreasonable in the present application, since the maintenance crew is likely to have some information regarding the unit's state during operation. This insight is used to perform maintenance so as to avoid a failure. Thus we are in practice faced with a situation of dependent competing risks between $W_1$ and $W_2$, and hence identifiability of marginal distributions requires additional assumptions.

Lindqvist, Støve and Langseth [49] suggested a model called the repair alert model to describe the joint behavior of the failure time $W_1$ and time $W_2$ of PM. This model is a special case of random signs censoring [15, 16] under which the marginal distribution of $W_1$ is always identifiable. Recall that $W_2$ is said to be a random signs censoring of $W_1$ if the event $\{W_2 < W_1\}$ is stochastically independent of $W_1$, that is, if the event of having a PM before failure is not influenced by the time $W_1$ at which the system fails or would have failed without PM. The idea is that the system emits some kind of signal before failure and that this signal is discovered with a probability which does not depend on the age of the system. The repair alert model extends this idea by introducing a so-called repair alert function which describes the "alertness" of the maintenance crew as a function of time.

Another possibility to obtain identifiability of the distributions of $W_1$ and $W_2$ would be to use the result of Zheng and Klein [64], which shows identifiability of marginal distributions when the dependence is given by a known copula.

Return now to the general case. Suppose that the system is repaired after failure and then put into operation, then may fail again and so on. This leads to a marked event process as described in Section 2 with marks in $\mathcal{J} = \{1, 2, \ldots, n\}$, so that the mark at each event time is the number of the failing component (or more generally the type of event).

The properties of this process depend on the repair strategy. Several classes of interesting models can be described in terms of a generalization of the virtual age concept introduced in Section 3.2, as discussed in the next subsection.

### 6.1 Virtual Age Models with Several Types of Events

Recall from Section 3.2 that the class of virtual age models generalizes the perfect repair and minimal repair models, and that the approach more generally leads to a large class of models. The main inputs are a hazard function $z(\cdot)$, which is thought of as the hazard function of a new unit, and a virtual age process which is a stochastic process which depends on the actual repair actions performed.

Several generalizations of the standard imperfect repair models are found in the literature. Shaked and Shanthikumar [60] suggested a multicomponent imperfect repair model with components that have dependent life-lengths. Langseth and Lindqvist [38] suggested a model which involves imperfect maintenance and repair in the case of several components and several failure causes. In a recent paper, Doyen and Gaudoin [24] developed the ideas further by presenting a general point process framework for modeling imperfect repair by a competing risks situation between failure and PM. Bedford and Lindqvist [8] considered a series system of $n$ repairable components where only the failing component is repaired at failures.

Inspired by the mentioned approaches, we suggest in this section a generalization of the imperfect repair models to the case where there is more than one type of event and where the virtual age process is multidimensional.

We let the first part of a virtual age model for $n$ components be given by a vector process $A(t) = (A_1(t), \ldots, A_n(t))$ that contains the virtual ages of the $n$ components at time $t$. The crucial assumption is that $A(t) = (A_1(t), \ldots, A_n(t)) \in \mathcal{F}_{t-}$, which means that the component ages are functions of the history up to time $t$.

As for the case with $n = 1$ in Section 3.2, it is assumed that the $A_j(t)$ increase linearly with time between events, and may jump only at event times. We define $v_j(i)$ to be the virtual age of component $j$ immediately after the $i$th event. The virtual age process for component $j$ is therefore defined by

$$A_j(t) = v_j(N(t-)) + t - T_{N(t-)}.$$

The second part of a virtual age model in the case $n = 1$ consists of the hazard function $z(\cdot)$. For general $n$ we replace this by functions $\nu_j(v_1, \ldots, v_n)$ for $v_1, v_2, \ldots, v_n \geq 0$, such that the conditional intensity of type $j$ events, given the history $\mathcal{F}_{t-}$, is

$$\gamma_j(t) = \nu_j(A_1(t), \ldots, A_n(t)).$$

Thus $\nu_j(v_1, \ldots, v_n)$ is the intensity of an event of type $j$ when the component ages are $v_1, \ldots, v_n$, respectively. The conditional intensity thus depends on the history only through the virtual ages of the components.

The family $\{\nu_j(v_1, \ldots, v_n) : v_1, v_2, \ldots, v_n \geq 0\}$ describes the failure mechanisms of the components and the dependence between them in terms of the ages of all the components. The basic statistical inference problem therefore consists of estimating these functions from field data. The case $n = 1$ has already been discussed in Section 3.2, but we shall see that identifiability problems can occur when $n > 1$.

### 6.2 Repair Models and their Virtual Age Processes

Most of the virtual age processes considered for the case $n = 1$ can be generalized to the present case of several event types. There are, however, often several ways to do this. Some examples are given below. Additional examples include generalizations of Kijima's [34] models, which may be plausible in applications.

6.2.1 *Perfect repair of complete system.* Suppose that all the components are repaired to as good as new at each failure of the system. In this case we have $v_j(i) = 0$ for all $j$ and $i$, and hence $A_j(t) = t - T_{N(t-)}$ for all $j$. It follows that we can only identify the "diagonal" values $\nu_j(t, \ldots, t)$ of the functions $\nu_j$. As noted in Section 6.3, these are given by the type-specific hazards defined in (14) for the nonrepairable competing risks case. This is not surprising in view of the fact that the present case of perfect repair essentially corresponds to observation of i.i.d. realizations of the nonrepairable competing risks situation.

6.2.2 *Minimal repair of complete system.* In the given setting a minimal repair will mean that following an event, the process is restarted in the same state as was experienced immediately before the event. In mathematical terms, this implies that $v_j(i) = T_i$ for all $i, j$ and hence that $A_j(t) = t$ for all $j$. Note that the complete set of functions $\nu_j$ is again not identifiable. Moreover, for a single component it is well known that minimal repair results in a failure time process which is a NHPP. In the present case of several components which are minimally repaired, it follows similarly that the failure processes of the individual components are independent NHPPs with the intensity for component $j$ given by $\nu_j(t, \ldots, t)$, which as already noted equals the type-specific hazard (14).

6.2.3 *A partial repair model.* Bedford and Lindqvist [8] suggested a partial repair model for the $n$ component case. The virtual age process is defined by letting $A_j(t) =$ time since last event of type $j$. Equivalently, the process could be defined by

$$v_j(i) = \begin{cases} 0, & \text{if } J_i = j, \\ v_j(i-1) + X_i, & \text{if } J_i \neq j. \end{cases}$$

Thus, the age of the failing component is reset to 0 at failures, whereas the ages of the other components are unchanged. The authors considered a single realization of the process, with the main result being that under reasonable conditions pertaining to ergodicity, the functions $\nu_j(v_1,\ldots,v_n)$ are identifiable. The intuitive idea of their proof is that the ages $v_1,\ldots,v_n$ will mix in such a manner that the complete set of $\nu_j(v_1,\ldots,v_n)$ can be identified.

6.2.4 *Age reduction models.* Doyen and Gaudoin [24] considered a single component or system and two types of events: $C_1$ = failure, $C_2$ = PM. In their basic model the virtual ages of the two types of events are equal: $A_1(t) = A_2(t) = A(t)$. They indicated, however, that this restriction is not necessary. Various choices of virtual age processes were considered. In particular they considered age reduction models that generalize these mentioned at the end of Section 3.2. More precisely, assume that there are given age reduction factors $0 < \rho_1, \rho_2 < 1$ for the two types of events. The virtual age immediately after the $i$th repair is then

$$v(i) = (1 - \rho_{J_i})(v(i-1) + X_i),$$

which means that the virtual age immediately before the $i$th failure, $v(i-1) + X_i$, is reduced due to repair by the factor $1 - \rho_{J_i}$. Alternatively, if only the additional age $X_i$ is reduced by the repair, it could be assumed that $v(i) = v(i-1) + (1 - \rho_{J_i})X_i$.

### 6.3 Modeling the Intensity Functions $\nu_j$

In principle the functions $\nu_j(v_1,\ldots,v_n)$ could be any functions of the component ages. Bedford and Lindqvist [8] motivated these functions by writing, for $j = 1,\ldots,n$,

(15) $\quad \nu_j(v_1,\ldots,v_n) = \lambda_j(v_j) + \lambda_{j*}(v_1,\ldots,v_n)$

with the convention that $\lambda_{j*}(v_1,\ldots,v_n) = 0$ when all the component ages except the $j$th are 0, so as to have uniqueness. Then $\lambda_j(v_j)$ is thought of as the intensity of component $j$ when working alone or together with only new components, while $\lambda_{j*}(v_1,\ldots,v_n)$ is the additional failure intensity imposed on component $j$ caused by the other components when they are not all new. Note that any functions of $v_1,\ldots,v_n$ can be represented this way, by allowing the $\lambda_{j*}$ to be negative as well as positive.

Langseth and Lindqvist [38] and Doyen and Gaudoin [24] extended the competing risks situation between failure and PM, as described at the beginning of the present section, and suggested how to define suitable functions $\nu_j$. The main ideas of these approaches can be described for general $n$ as follows. Starting from a state where the component ages are, respectively, $v_1,\ldots,v_n$, let the time to next event be governed by the competing risks situation between the random variables $W_1^*,\ldots,W_n^*$ with distribution equal to the conditional distribution of $W_1 - v_1,\ldots,W_n - v_n$ given $W_1 > v_1,\ldots,W_n > v_n$, where the $W_i$ are defined in the nonrepairable case described at the beginning of the section. It is then rather straightforward to show that this implies

(16) $\quad \nu_j(v_1,\ldots,v_n) = \dfrac{-\partial_j R(v_1,\ldots,v_n)}{R(v_1,\ldots,v_n)},$

where $R(v_1,\ldots,v_n) = P(W_1 > v_1,\ldots,W_n > v_n)$ is the joint survival function of the $W_i$, and $\partial_j$ means the partial derivative with respect to the $j$th entry in $R$. Note that this generalizes the usual hazard rate in the case $n = 1$ considered in Section 3.2. Further, we have $\nu_j(t,t,\ldots,t) = h_j(t)$, where the latter is the type-specific hazard rate given in (14).

A final remark on the suggested construction of the functions $\nu_j$ is due. It was demonstrated by Bedford and Lindqvist [8] that, even in the case with $n = 2$, it is not always possible to derive a general set of functions $\nu_j(v_1,\ldots,v_n)$ from a single joint survival distribution as in (16). A simple counterexample was given in [8]. Thus for generality one should stick to completely general representations like (15).

## 7. CONCLUDING REMARKS

In the present paper we have reviewed some main approaches for the analysis of data from repairable systems. To a large extent the emphasis has been on describing the underlying principles and structures of common models. Essential features of such models correspond to the three dimensions of the model cube in Figure 3: renewal property, time trend and heterogeneity. The presentation places less emphasis on statistical inference than on modeling. However, it has been an intention to show how likelihood functions are obtained for the different models. It is also indicated how covariates can be included in the models and the corresponding likelihood functions. While the derived likelihood functions can be used in a rather straightforward manner in parametric statistical inference, there turn out to be several challenging problems connected to nonparametric estimation in some of the models.

Two main types of models with rather simple and transparent basic structures have been considered. These are the virtual age type models and the TRP type models. The former type combines two basic ingredients: a hazard rate $z(\cdot)$ of a new system together with a particular repair strategy which governs the virtual age process $A(t)$. The renewal dimension is taken care of by the virtual age process, while trend is determined by the distribution of a new system. For the $\text{TRP}(F, \lambda(\cdot))$, the renewal dimension corresponds to the renewal distribution $F$, while the trend is explicitly given by the trend function $\lambda(\cdot)$. For both types of processes, heterogeneity can be included by multiplicative factors working on the intensities. A noticeable difference between the two types of models as regards statistical inference is that the virtual age type model usually requires that the virtual age process be observable. Such observations may, however, often be lacking in real data.

Many processes show some degree of clustering of failures. This may be due to various causes; see, for example, [17]. Several models have been suggested in the literature, a classical one being the Neyman and Scott [54] model. As pointed out by a referee, even the TRP model can pick up the clustering effect by allowing the renewal distribution to be a mixture with a substantial amount of probability near zero.

Peña [55] has reviewed a class of models suggested in [56]. These are virtual age models which include the possibility of heterogeneity between systems, time-dependent covariates, and for which in addition the conditional intensities may depend on the number of previous events. This last feature adds an interesting flexibility to the model. In particular it enables modeling of certain load sharing processes and software failure processes.

Certain systems, for example, alarm systems, are tested only at fixed times which are usually periodic. If the system is found in a failed state, then it is repaired or replaced. Thus repair is not done at the same time as the failure, and the situation is not covered by the methods considered in the paper. A simple model of this situation was suggested by Hokstad and Frøvig [30] and further studied and extended by Lindqvist and Amundrustad [47]. Consider a system which starts operation at time $t = 0$ and is tested at time epochs $\tau, 2\tau, 3\tau, \ldots$. When time is running between testing epochs, the state of the system is modeled by an absorbing Markov chain. Having thus defined the probabilistic behavior of the system state between testing, one needs to add to the model a specification of the repair policy. In [47] this is modeled in the form of a transition matrix on the state space of the Markov chain, which defines the possible changes of state and their probabilities following the repair actions.

In a given study there is usually a choice between several types of models. It is thus important to have tools for model checking and goodness-of-fit procedures. For model checking in parametric estimation of the HTRP model, we refer to [48], which used a type of Cox–Snell residuals together with plots using the TTT technique. The general underlying idea, which in principle can be used with all estimation methods considered in this paper, is that the process of integrated conditional intensities, $\int_0^{T_1} \gamma(t)\,dt, \int_0^{T_2} \gamma(t)\,dt, \ldots$, is HPP(1) [12]. In turn this gives rise to computable residual processes when estimates are inserted for parameters and distributions. The use of these processes in model checking is demonstrated for three different data sets in [48]. Typically, one would check (i) the distribution of the residuals with respect to departures from the unit exponential distribution, (ii) the possible presence of time trends in residuals within each system and (iii) the possible presence of autocorrelation in times between events in the residual processes.

## ACKNOWLEDGMENTS

The author is grateful to Dr. Sallie Keller-McNulty and Dr. Alyson Wilson for the invitation to contribute to this special issue of *Statistical Science*. The paper was prepared while the author was working in an international research group studying event history analysis at the Centre for Advanced Study at the Norwegian Academy of Science and Letters in Oslo during the academic year 2005/2006. The author is thankful for the pleasant hospitality extended to him. Discussions with other members of the group, in particular Odd Aalen, Per Kragh Andersen and Ørnulf Borgan, have been very helpful and are greatly appreciated. Thanks are also due the author's former PhD students Georg Elvebakk, Jan Terje Kvaløy and Helge Langseth for long time collaborations and important contributions to the theory presented here. The author finally thanks two anonymous referees for careful reading of the manuscript and for their important and constructive comments, which led to a much improved paper.